\documentclass[prd,nofootinbib,floats,superscriptaddress,eqsecnum,tightenlines,preprintnumbers,11pt]{revtex4-1}

\usepackage{mathtools}
\usepackage{amssymb}
\usepackage{dsfont}
\usepackage{lipsum}
\usepackage{mathtools}
\usepackage{xcolor}
\usepackage[colorlinks=true,linkcolor=blue,citecolor=magenta,linktocpage=true]{hyperref}
\usepackage{physics}
\usepackage{csquotes}
\usepackage{mathtools, stmaryrd}
\usepackage{bm}
\usepackage{array}
\usepackage{multirow}
\usepackage{tensor}
\usepackage{interval}

\newcolumntype{M}[1]{>{\centering\arraybackslash$\displaystyle}p{#1}<{$}}

\usepackage{caption}
\usepackage{subcaption}

\usepackage[makeroom]{cancel}

\def\be{\begin{equation}}
	\def\ee{\end{equation}}
\def\bea{\begin{eqnarray}}
	\def\eea{\end{eqnarray}}
\def\beq{\begin{eqnarray}}
	\def\eeq{\end{eqnarray}}
\def\bas{\begin{subequations}\begin{eqnarray}}
		\def\eas{\end{eqnarray}\end{subequations}}

\def\rd{\mathrm{d}}
\def\veps{\varepsilon}

\newcommand{\cT}{{\mathcal T}}

\DeclareMathOperator{\arctanh}{arctanh}
\newcommand{\D}{\partial}
\newcommand{\Db}{\bar{\partial}}
\newcommand{\Dz}{\mathcal{D}_0}
\newcommand{\bn}{{\bm{n}}}

\begin{document}

	\title{Algebraically special perturbations of the Kerr black hole: \\ A metric formulation}
	
	\author{Jibril Ben Achour}
	\affiliation{Arnold Sommerfeld Center for Theoretical Physics, Munich, Germany}
	\affiliation{Munich Center for Quantum Science and Technology, Munich, Germany}
	\affiliation{Univ de Lyon, ENS de Lyon, Laboratoire de Physique, CNRS UMR 5672, Lyon 69007, France}
	\author{Clara Montagnon}
	\affiliation{Univ de Lyon, ENS de Lyon, Laboratoire de Physique, CNRS UMR 5672, Lyon 69007, France}
	\author{Hugo Roussille}
	\affiliation{Univ de Lyon, ENS de Lyon, Laboratoire de Physique, CNRS UMR 5672, Lyon 69007, France}


	\begin{abstract}
Perturbations of the Kerr black hole are notoriously difficult to describe in the metric formalism and are usually studied in terms of perturbations of the Weyl scalars. In this work, we focus on the algebraically special linear perturbations (ASLP) of the Kerr geometry and show how one can describe this subsector of the perturbations solely using the metric formulation. To that end, we consider the most general twisting algebraically special solution space of vacuum General Relativity. By linearizing around the Kerr solution, we obtain two coupled partial differential wave equations describing the dynamics of the Kerr ASLP. We provide an algorithm to solve them analytically in the small spin approximation up to third order, providing the first exact solution of this kind in the metric formulation. Then, we use this framework to study the stationary zero modes of the Kerr geometry. We present the exact analytical form of the shifts in mass and spin together with the required change of coordinates needed to identify them. Finally, we also provide for the first time closed expressions for the solution-generating perturbations generating the NUT and acceleration charges, thus deforming the Kerr solution to the linearized Kerr-NUT and spinning C-metric. These results provide a first concrete and rare example of perturbations of the Kerr black hole which can be treated entirely in the metric formulation. They can serve as a useful testbed to search for hidden symmetries of the Kerr perturbations.
	\end{abstract}

	\maketitle
	\tableofcontents
	
	\newpage
	
	\section*{Introduction}
	
	Black hole perturbation theory stands as a cornerstone formalism to study the late time ringdown phase following a binary black hole merger, compute the quasi-normal modes spectrum, evaluate the energy fluxes and characterize the properties of the perturbed geometry~\cite{Berti:2025hly, Martel:2005ir, Berti:2004md, Berti:2009kk, Berti:2015itd, Franchini:2023eda}. The study of linear perturbations around a Schwarzschild black hole was initiated by the seminal work of Regge and Wheeler in~\cite{Regge:1957td}. As shown there, the apparently complicated set of ten coupled partial differential equations (PDE) descending from the linearization of the Einstein equations 
	on the Schwarzschild geometry can be drastically simplified using both background symmetries and gauge freedom. Indeed, the spherical symmetry of the background geometry allows one to organize the perturbations w.r.t their properties under parity transformations, giving rise to the axial and polar sectors. Decomposing the perturbation profiles into spherical harmonics labelled by their angular momentum and magnetic numbers $(\ell,m)$, one effectively decouples axial and polar perturbations,  reducing the problem to a set of coupled PDE for each sector. Then, one can choose a well adapted gauge, i.e. the Regge-Wheeler gauge fixing, to set four of the ten components of $h_{\mu\nu}$ to zero. Finally, upon using the time-translation symmetry to decompose the remaining fields into modes, one finally obtains two decoupled ordinary differential equations (ODE) for the radial profile of the (axial and polar) perturbations. Therefore, the initial ten coupled PDE can be reduced to two decoupled radial ODE of the Schrodinger-type. This state of affair is no longer true when considering the rotating Kerr black hole as the background geometry.
	
	First, the Kerr geometry breaks spherical symmetry, preventing one to organize the perturbations into axial and polar sectors. Second, the linearized Einstein equations on the Kerr black hole do not decouple and lead to a highly complicated system of PDE. So far, no adapted gauge choice generalizing the Regge-Wheeler gauge has been found for the metric perturbations around Kerr. As a result, this prevents one to directly analyze the Kerr perturbations in the metric formulation. Instead, it was shown by Teukolsky that the problem can be reformulated in terms of perturbations of the Weyl tensor $C_{\mu\nu\rho\sigma}$. Using the Newman-Penrose formalism and appropriately choosing a null frame, he proved that the Weyl perturbations surprisingly satisfy a set of decoupled and separable waves equations known today as the Teukolsky master equations~\cite{Teukolsky:1973ha}. Quite remarkably, he further showed that the dynamics of the perturbations of any spin-$s$ field on the Kerr black hole can be collectively captured within the same master equations, the spin-$2$ gravitational perturbations corresponding to the two gauge-invariant perturbations of the Weyl scalars $(\delta \Psi_0, \delta \Psi_4)$. This beautiful result opened the road to the analytical study of the dynamics of the perturbations of the Kerr black hole, computing its quasi-normal modes spectrum and the gravitational waveform emitted during its relaxation to equilibrium~\cite{Bardeen:1972fi, Press:1973zz, Teukolsky:1974yv}.
	
	Yet, while the Teukolsky master equations can be solved analytically and their solutions can be expressed in terms of suitable special functions, i.e. the Heun functions, a direct computation of the associated metric perturbations turns out to be highly non-trivial. The task of providing a metric reconstruction algorithm to obtain the perturbation of the metric $h_{\mu\nu}$ given the solutions  $(\delta \Psi_0, \delta \Psi_4)$ of the Teukolsky equations has attracted considerable efforts~\cite{Chrzanowski:1975wv, Wald:1978vm, Kegeles:1979an, Ori:2002uv, Berens:2024czo}. Indeed, this is crucial in order to analyze the properties of the perturbed geometry as well as to investigate the non-linear perturbative regime, a topic which has attracted important efforts recently \cite{London:2014cma, Cheung:2022rbm, Mitman:2022qdl, Ma:2022wpv, Lagos:2022otp, Bhagwat:2023fid, Kehagias:2023ctr, Khera:2023oyf, Redondo-Yuste:2023seq, Perrone:2023jzq, Bucciotti:2023ets, Cheung:2023vki, Qiu:2023lwo, Zhu:2024rej, Yi:2024elj, Bucciotti:2024zyp, Bourg:2024jme, DeLuca:2023mio, Kehagias:2024sgh, Kehagias:2025ntm, Kehagias:2025xzm, Kehagias:2025tqi, Perrone:2025zhy, Ma:2024qcv,Khera:2024bjs}. Starting from the solutions for $(\delta \Psi_0, \delta \Psi_4)$, one has to determine auxiliary potentials known as the Hertz potentials, which require a inversion operation which has to be carried out mode by mode, and finally use suitable differential operators to reconstruct metric perturbations. We refer the reader to the recent work~\cite{Berens:2024czo} for a detailed account on this procedure. So far, this reconstruction algorithm is the only known procedure to access to the metric perturbations of the Kerr black hole. 
	
	In view of this state of affairs, it is interesting to wonder if one can identify a subsector of the Kerr perturbations which can be treated solely within the metric formulation, i.e. without resorting to the Teukolsky potentials and master equations. It turns out that algebraically special perturbations (ASP) of the Kerr black hole provide such a rare and interesting example. The goal of this work is to provide a detailed study of the linear ASP of the Kerr geometry. To that end, we shall extend our previous work on the Schwarzschild ASP~\cite{BenAchour:2024skv} using the strategy initially introduced by Couch and Newman in~\cite{Couch:1973zc} and revisited by Qi and Schutz in~\cite{Qi:1993ey}. Let us point that besides providing a rare example allowing for a purely metric-based treatment, the study of the ASP is also important for at least three reasons. 
	
	In the spherically symmetric case, the Schwarzschild ASP have played a key role in the seminal investigation by Chandrasekhar leading to the proof of the isospectrality of the axial and polar sectors~\cite{Chandrasekhar1984}. In that sense, they can serve as non-trivial but simple enough exact solutions to search for hidden symmetries of black hole perturbations.  See~\cite{ISO, Glampedakis:2017rar, Yurov:2018ynn, Lenzi:2021njy, Lenzi:2023inn} for details on the isospectrality and the role of the ASP in identifying this property.
	
	The second interesting outcome when studying ASP is that by definition, they contain all the stationary zero modes of the background geometry\footnote{In higher dimensions, i.e. $d>4$, it has been shown that the ASP only describe zero modes of the background geometry~\cite{Dias:2013hn}, in contrast with the four dimensional counterpart where one exhibits propagating ASP modes.}. These zero modes can be understood as specific perturbations generating non-trivial charge, i.e. large gauge transformations, connecting thus the background solution to another inequivalent stationary solution of Einstein equations. For instance, the Kerr solution belongs to the more general Plebianski-Demianski familly of exact electro-vacuum solutions of Petrov type D~\cite{Plebanski:1976gy, Griffiths:2005qp}. Focusing on the vacuum case as we should do here, this family contains the Kerr-NUT and the spinning C-metric which can be understood as a deformation of the Kerr geometry with respectively a non-vanishing NUT and acceleration charge. Therefore, there should exist zero modes of the perturbations which map the Kerr metric to the linearized versions of the Kerr-NUT and spinning C-metric~\cite{Bicak:1999sa, Hong:2004dm}, i.e. considering the NUT and acceleration charge as small parameters. To our knowledge, their explicit expressions has not been provided yet in the literature.  As we shall see, our metric-based approach of the ASP allows us to provide their exact expressions in a closed form. This is to be contrasted with the treatment of the Kerr ASP presented by Wald in~\cite{Wald:1973wwa} and revisited by Chandraskehar in~\cite{Chandrasekhar:1984mgh} based on the NP formalism.
	
	Another interesting feature which makes the ASP interesting on their own right is that there exist modes satisfying the boundary conditions of either the quasi-normal modes (QNM), the totally transmitted modes (TTM) or none of the two, which exhibit a negative imaginary frequency which is numerically infinitely close to the analytic ASP frequency (see~\cite{MaassenvandenBrink:2000iwh} for a detailed study of these purely imaginary modes for the Schwarzschild black hole). It is remarkable but also mysterious that perturbations satisfying very different boundary conditions turn out to have very close characteristic frequencies. On one hand, the algebraically special QNM/TTM stand as a subset of the spectrum of well defined perturbations with compact support which allow the Schwarzschild geometry to return to equilibrium. On the other hand, the ASP studied in~\cite{Couch:1973zc, Qi:1993ey, BenAchour:2024skv} are selected solely by the requirement of preserving the algebraic specialness of the perturbed geometry, leading generically to pathological perturbation profiles which blow up at the horizon or at infinity. Purely imaginary frequency modes of the Kerr perturbations have been analyzed in detail in~\cite{Cook:2016ngj}, identifying different families of modes whose frequencies are also numerically infinitely close to the ASP ones (see~\cite{Berti:2003jh, Cook:2016ngj, Cook:2016fge, Cook:2016fge, Hod:2013fea} for a discussion on the purely imaginary QNM of the Kerr black hole). Therefore, it appears useful to further explore the description of the Kerr ASP and provide a metric-based description, a task which has not been tackled yet in the literature.
	
	This work is organized as follows. In Section~\ref{I}, we present the form of the Teukolsky master equations in term of the Weyl scalars and we provide a short review of the work of Wald on the Kerr ASP within that formalism. In Section~\ref{II}, we present the solution space describing the most general algebraically special twisting vacuum solutions of General Relativity. Using this solution space, we show how the linearization around the Kerr solution allows one to derive two coupled dynamical equations given by Eq~(\ref{eq:pert-axial}) and Eq~(\ref{eq:pert-polar}) for the linear Kerr ASP. In Section~\ref{III}, we show how these coupled PDE can be analytically solved in the small spin regime up to an arbitrary order.  Finally, Section~\ref{IV} is devoted to a detailed description of the Kerr zero modes. We first describe the large-gauge perturbations which generate the quadratic corrections to the mass and spin. Then, we present the solution-generating perturbations (\ref{NUT1} - \ref{NUT2}) and (\ref{C1} - \ref{C2}) generating the linearized Kerr-NUT and spinning C-metric, together with the non-trivial change of coordinates needed to identify these representants of the Plebianski-Demianski family of exact solutions.  We conclude in Section~\ref{V} by a discussion on the perspectives opened by this work.

	\section{Algebraically special modes from the Teukolsky equation}
	
	\label{I}
	
	In this section, we review the Teukolsky master equations and the procedure followed by Wald to characterize the Kerr ASP in~\cite{Wald:1973wwa}. This will reveal useful to contrast with our metric-based approach.
	
	\subsection{Teukolsky master equations}
	
	As a starting point, let us review the motivation and the derivation of the Teukolsky equations for the Kerr perturbations. Consider the Kerr background solution with line element in Boyer-Lindquist coordinates given by
	\begin{equation}
		\dd{s}^2 = - \Big(1 - \frac{2Mr}{\rho \bar{\rho}}\Big)\dd{t}^2 + \frac{\rho \bar{\rho}}{\Delta} \dd{r}^2 +\rho \bar{\rho} \dd{\theta}^2 + \Big(r^2 + a^2 + \frac{2M r}{\rho \bar{\rho}} a^2 \sin^2\theta\Big) \sin^2\theta \dd{\varphi}^2 - \frac{4aMr}{\rho \bar{\rho}} \sin^2\theta \dd{t} \dd{\varphi}\,,
		\label{kerr_metric}
	\end{equation}
	where the functions $\Delta$ and $\rho$ are given by
	\begin{equation}
		\Delta = r^2 - 2Mr + a^2 \qq{and} \rho = r + ia\cos\theta \,.
	\end{equation}
	One could consider perturbing the Kerr metric $g^\mathrm{K}_{\mu\nu}$, i.e. consider $g_{\mu\nu} = g^\mathrm{K}_{\mu\nu} + h_{\mu\nu}$ and study the dynamics of the metric perturbations $h_{\mu\nu}$, which satisfy the linearized Einstein equations given by
	\begin{align}
		& \Box^\mathrm{K} h_{\mu\nu} + \nabla^\mathrm{K}_{\mu} \nabla^\mathrm{K}_{\nu} h - g^\mathrm{K}_{\mu\nu} \left(  \Box^\mathrm{K} h -\nabla^\mathrm{K}_{\mu} \nabla^\mathrm{K}_{\nu} h^{\mu\nu} \right) + \nabla^\mathrm{K}_{\mu} \nabla^\mathrm{K}_{\nu} h =0 \,,
	\end{align}
	where the subscript ${}^\mathrm{K}$ corresponds to evaluation on the Kerr background. However, this set of ten coupled PDEs turns out to be highly complicated to study. In contrast to the perturbations on a spherically symmetric background, no one has found a suitable gauge fixing allowing to simplify and organize the physical degrees of freedom in term of the components of the metric perturbations $h_{\mu\nu}$. In particular, to date, it is not known how to turn the above system of equations into a decoupled and separable one for the Kerr metric perturbations.
	
	Remarkably, it was shown by Teukolsky that instead of using the metric perturbations $h_{\mu\nu}$, the problem can be recast in a surprisingly simple and tractable form when expressed in terms of the perturbations of the Weyl tensor whose dynamics can be studied using the Newman-Penrose formalism~\cite{Teukolsky:1973ha}. Concretely, as the Kerr solution belongs to the Petrov type D family, there exists a null tetrad $(\ell^\mu, n^\mu, m^\mu, \bar{m}^\mu)$ given by
	\begin{subequations}
		\begin{align}
			\ell^\mu \partial_\mu &= \frac{1}{\Delta} \big[(r^2 + a^2)\partial_t + \Delta \partial_r  + a \partial_\varphi\big]\,,\\
			n^\mu \partial_\mu &= \frac{1}{2 \rho \bar{\rho}} \big[(r^2 + a^2)\partial_t - \Delta \partial_r  + a \partial_\varphi\big] \,,\\
			m^\mu \partial_\mu &= \frac{1}{\rho\sqrt{2}} \big(i a \sin\theta \partial_t + \partial_\theta + \frac{i}{\sin\theta} \partial_\varphi\big)\,,
		\end{align}
	\end{subequations}
	for which four of the five Weyl scalars vanish, i.e. 
	\begin{equation}
		\Psi_0 = \Psi_1 = \Psi_3 = \Psi_4 = 0 \,,
	\end{equation}
	while $\Psi_2$ is given by
	\begin{equation}
		\Psi_2 = - \frac{M}{\rho^3}\,.
	\end{equation}
	This tetrad corresponds to the principal null directions of the Kerr geometry.
	Following Teukolsky, one can consider the perturbations of the Weyl scalars, i.e. $\delta \Psi_i$. Linearizing the Newman-Penrose equations (and in particular the Bianchi identities which translate into evolution equations for the $\delta \Psi_i$), one can perform a suitable sequence of Lorentz transformations on the initial PNDs to show that the perturbations can be fully encoded in the two gauge invariants fields $\delta \Psi_0$ and $\delta \Psi_4$ which correspond respectively to the ingoing and outgoing radiative parts of the Weyl tensor. Concretely, introducing the two auxiliary fields
	\be
	\psi^{(+2)} = \delta \Psi_4 \,, \qquad \psi^{(-2)} = (r^2 - ia\cos^2{\theta}) \; \delta \Psi_0 \,,
	\ee
	one can show that both satisfy a decoupled wave equation of the form $\cT^{s} \psi^{s} =0$ where the Teukolsky operator  $\cT^{s}$ reads
	\begin{equation}
		\begin{aligned}
			\cT^{s} & = \left( \frac{(r^2+a^2)^2}{\Delta} - a^2 \sin^2{\theta}\right) \partial^2_t - 2s \left( \frac{(r^2-a^2)^2}{\Delta} -r - i a \cos{\theta}\right) \partial_t + \frac{4a M r}{\Delta} \partial_t \partial_r \\
			&  - \Delta^{-s} \partial_r (\Delta^{s+1}\partial_r ) - \frac{1}{\sin{\theta}} \partial_{\theta} (\sin{\theta} \partial_{\theta}) + \left( \frac{a^2}{\Delta} - \frac{1}{\sin^2{\theta}}\right) \partial^2_{\phi}  + s \left( \frac{s}{\tan^2{\theta}} -1\right) \,.
		\end{aligned}
	\end{equation}
	This is the well-known Teukolsky master equation. Using the stationary and the azimutal Killing symmetries of the Kerr geometry, one can decompose the field $\psi^{s}$ as
	\begin{align}
		\psi^{s}(t,r,\theta, \phi) = \int \rd \omega \sum^{\infty}_{\ell=|s|} \sum^{+\ell}_{m=-\ell} e^{-i\omega t + i m \phi} S^s_{\omega \ell m} (\theta)R^s_{\omega \ell m}(r) \,.
	\end{align}
	This ansatz separates the master equation into two second order ODE of the form
	\begin{subequations}
		\begin{align}
			\label{ekk1}
			\left[ \frac{1}{\sin{\theta}} \frac{\rd }{\rd \theta} \left(\sin{\theta} \frac{\rd }{\rd \theta} \right) + a^2 \omega^2 \cos^2{\theta} - \frac{(m+s\cos{\theta})^2}{\sin^2{\theta}} - 2 s a \omega \cos{\theta} + s + \mathcal{A}_{\omega \ell m} \right] S^s_{\omega \ell m} (\theta) & =0 \\
			\label{ekk2}
			\left[ \frac{1}{\Delta^s} \frac{\rd }{\rd r} \left(\Delta^{s+1} \frac{\rd }{\rd r} \right) + \frac{K^2 - 2is (r-M) K}{\Delta} + 4 i s \omega r - \lambda^s_{\omega \ell m} \right] R^s_{\omega \ell m} (r) & =0 
		\end{align}
	\end{subequations}
	where $\lambda^s_{\omega \ell m}$ is a separation constant and the two constants $(\mathcal{A},K)$ are given by
	\be
	\label{con}
	\mathcal{A}_{\omega \ell m}= 2 a m \omega - (a\omega)^2 + \lambda^s_{\omega \ell m} \,, \qquad K = (r^2+a^2) \omega - a m \,.
	\ee
	Notice that $\lambda^s_{\omega \ell m}$ is only known numerically, although an analytical expression is known in the small spin regime. 
	The angular equation defines the so called spin-weighted spheroidal harmonics $S^s_{\omega \ell m} $ which are the adapted angular functions needed to decompose the polar profile of the Weyl perturbations on the rotating Kerr geometry~\cite{Goldberg:1966uu}.
	
	Both the angular and radial equations can be recast into an equation satisfied by confluent Heun functions (or related functions), allowing one to represent the solutions in term of these special functions. We refer the interested reader to~\cite{Borissov:2009bj} for a detailed discussion on this point. We are now ready to review the characterization of the algebraically special perturbations of the Kerr black hole studied by Wald.

	\subsection{The algebraically special sector: Wald's approach}
	\label{Wald}
	
	General perturbations generically break the algebraic specialness of a given background geometry~\cite{Araneda:2015gsa}. Yet, it is always possible to identify a subset of perturbations which lead to an algebraically special perturbed solution, i.e. the so called algebraically special perturbations.  Whether this subset contains non-trivial propagating modes depends on the spacetime dimension and the background solution one considers. For instance, it is known that in dimension greater than four, all ASP are pure zero modes, i.e. they do not contain any dynamical propagating modes~\cite{Dias:2013hn}. This is to be contrasted with the ASP of the four dimensional Schwarzschild black hole which were initially studied by Chandrasekhar in~\cite{Chandrasekhar1984}, and later on by Couch and Newman in~\cite{Couch:1973zc}. Indeed, the Schwarzschild black hole admits a whole family of non-trivial ASP modes with purely imaginary frequency of the form
	\begin{equation}
		\omega_{\ell} =  \pm \frac{(\ell - 1)\ell(\ell + 1)(\ell + 2)}{12M} i \,,
		\label{schwa_special_modes}
	\end{equation}
	These dynamical modes do not satisfy the boundary conditions for well-behaved perturbations as they blow up either at the horizon or at infinity and are therefore usually ignored. Nevertheless, taking into account the Schwarzschild background dressed with these linear ASP stands as a \textit{bona fide} radiative Petrov type II exact solution of the (linearized) Einstein equations. As first shown in~\cite{Couch:1973zc}, these exact solutions can also be obtained by linearizing the solution space of algebraically special radiative geometries such as the vacuum Robinson-Trautman solution space, or as we shall see, the solution space of the vacuum twisting AS geometries.
	
	As expected, due to its inherent complexity, the treatment of the ASP perturbations of the Kerr geometry has received much less attention. To our knowledge, the first and only characterization of the Kerr ASP was discussed by Wald in~\cite{Wald:1973wwa}. The central motivation of this work was not directly related to the Kerr ASP but instead to demonstrate that for well behaved perturbations, each of the two perturbations $\delta \Psi_4$ and $\delta \Psi_0$ uniquely determine each other. In a modern form, this translates into the fact that these two perturbations are actually not independent, as can be seen from the well-known Teukolsky-Starobinski identities. 
	
	Wald first demonstrated that for well-behaved perturbations with $\delta \Psi_0 =0$, one automatically has $\delta\Psi_4 =0$. As it turns out, such perturbations are by construction algebraically special of Petrov type D. This subset of the Kerr perturbations contains four linearly independent solutions which correspond to the ones involving a change of mass, spin or a linearized NUT or acceleration charge. However, while Wald successfully identified each of these exact solutions using the form of the tetrad, he did not provide the expressions for the metric perturbations relating the background Kerr solution to any of the four representants of the Petrov type D solution space. As we shall see later, the metric-based approach adopted in the present work will allow us to fill this gap and provide explicit expressions for these metric perturbations.
	
\begin{figure}[!htb]
		\centering
		\includegraphics{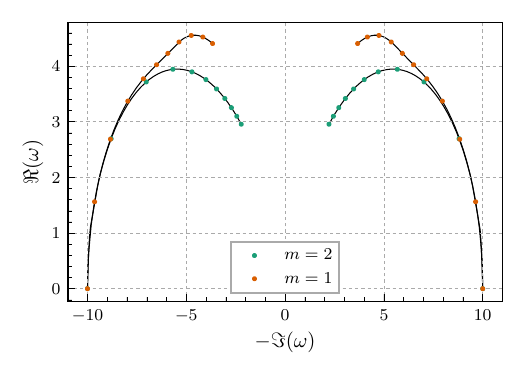}
		\caption{Solutions $\omega$ to equation~\eqref{eq:det-chandra} for $\ell = 3$ and different values of $m$. The mass $M$ is set to 1, and the spin $a$ spans the interval $\interval{0}{0.5}$, with a dot every 0.05.}
		\label{fig:plot-asms}
	\end{figure}

	Beyond this type D subset of zero modes, one has a whole family of other dynamical perturbations of Petrov type II for which $\delta \Psi_4 \neq 0$. Following Wald, these dynamical ASP can be investigated by first imposing $\delta \Psi_0=0$, which ensures that we are dealing with an AS perturbed geometry thanks to the linearized version of the Goldberg-Sachs theorem, and then use the simplified Newman-Penrose equations to solve the radial dependence of $\delta \Psi_4$. By comparing with the radial part of the Teukolsky equation for gravitational perturbations, he found that both are compatible provided the frequency of these dynamical perturbations satisfies the condition
	\begin{equation}
		\Lambda^2 (\Lambda - 2)^2 - 12 \Lambda (\Lambda - 2) \Gamma \omega - 8 \Lambda^2 \Gamma \omega + 36 (\Gamma^2 + 4M^2) \omega^2 - 96 \Lambda a^2 \omega^2 = 0 \,,
		\label{eq:det-chandra}
	\end{equation}
	where
	\begin{align}
		\Gamma &= 2 \omega a^2 - 2 a m \,, &\Lambda &= 2 a m \omega - a^2\omega^2 -  \mathcal{A}_{\omega \ell m}\,.
	\end{align}
	Here, $\mathcal{A}_{\omega \ell m}$ is the eigenvalue associated with the spin-weighted spheroidal harmonic given in~\eqref{con}. In the limit $a = 0$, the equation~\eqref{eq:det-chandra} defining the ASM reduces to
	\begin{equation}
		(\ell - 1)^2(\ell + 2)^2 \big[(\ell - 1)(\ell + 2) - 2\big]^2 + 144M^2 \omega^2 = 0 \,,
	\end{equation}
	for which the roots are given precisely by (\ref{schwa_special_modes}), recovering the well-known ASM modes of a Schwarzschild BH solution~\cite{Couch:1973zc,BenAchour:2024skv}. However, solving this equation for $a\neq 0$ to obtain the form of $\omega$ turns out to be much more involved than in the case of the Schwarzschild ASP. The key difference is that the constant $\mathcal{A}_{\omega \ell m}$ now depends on the frequency $\omega$. One possibility is to solve the equations numerically. In figure~\ref{fig:plot-asms}, we plot the numerical solution for $\omega$ for $\ell = 3$ and different values of $m$.  It is direct to see that in contrast with the Schwarzschild case, the frequency of the ASP is no longer purely imaginary. This implies that in general, the ASP are not purely decaying but also exhibit an oscillating behavior.
	
	Let us also point that while a resolution of (\ref{eq:det-chandra}) for the Kerr case is challenging, one can consider the slow rotation regime. Using an expansion for $a\omega \ll 1$ of the angular eigenvalue $\mathcal{A}_{\omega \ell m}$, one obtains
	\begin{equation}
		\mathcal{A}_{\omega \ell m}= \ell(\ell+1) - 2 - \frac{8ma\omega}{\ell(\ell+1)} + \frac{\alpha_{\ell m} (a\omega)^2}{\ell^3(\ell+1)^3[4\ell(\ell+1) - 3]} + \mathcal{O}(a^3\omega^3) \,,
	\end{equation}
	where
	\begin{equation}
		\alpha_{\ell m} = 2m^2 \big[ (\ell-4) \ell (\ell+1) (\ell+5) (\ell(\ell+1) - 4) +48 \big]- \ell^2 (\ell+1)^2 \big[ \ell(\ell+1)(2\ell(\ell+1) - 17) + 96 \big] \,.
	\end{equation}
	The roots of~\eqref{eq:det-chandra} can then be written explicitly up to the cubic order in spin and read
	\begin{equation}
		\omega^\pm = \pm i\frac{(\ell-1)\ell(\ell+1)(\ell+2)}{12M} + \frac{a m \ell(\ell+1)[\ell(\ell+1)-2]^2}{36M^2} \mp \frac{i a^2 \beta_{\ell m}}{432 M^3 [4\ell(\ell+1) - 3]} + \mathcal{O}(a^3) \,,
		\label{expansion_om}
	\end{equation}
	where the constant $\beta_{\ell m}$ is given by
	\begin{equation}
		\beta_{\ell m} = [\ell(\ell+1)-2]^3 \big[ m^2 [\ell(\ell+1) (21\ell(\ell+1) +2) + 12] + (\ell-3)\ell^2(\ell+1)^2(\ell+4) \big] \,.
	\end{equation}
	This provides an analytical expression for the frequency of the dynamical ASP modes propagating on the slowly rotating Kerr geometry.  
	
	At this stage, let us emphasize that Wald's approach relies on the use of the Newman-Penrose formalism and the Teukolsky master equation to derive the equation~\eqref{eq:det-chandra} dictating the ASP frequency. Yet, no description of the metric perturbations profiles was provided in this work. In the following, we will show a straightforward path to obtain such a metric description of the Kerr ASP.
	
	\section{General solution space for algebraically special black hole perturbations}
	
	\label{II}
	
	In this section, we review the subsector of the full Einstein equations which describes the most general twisting vacuum algebraically special geometries. This solution space will be central to describe all the algebraically special perturbations of the Kerr black hole.
	
	\subsection{Petrov type II twisting vacuum radiative geometries}
	
	The construction of the solution space of the most general algebraically special twisting solutions and the associated reduction of the Einstein equations were investigated in the sixties~\cite{Kerr:1963ud, Debney:1969zz, Wain1974} (see~\cite{BernardideFreitas:2015lyp} for related investigations in higher dimensions). In this section, we follow closely~\cite{Stephani:2003tm} to present this solution space. Consider the line element given by
	\begin{equation}
		\dd{s}^2 = -2 \omega^l \omega^n + 2 \omega^m \omega^{\bar{m}} \,,
		\label{eq:metric}
	\end{equation}
	where the tetrad $\omega $ is
	given by
	\begin{equation}
		\omega^l = \dd{u} + L \dd{z} + \bar{L} \dd{\bar{z}} \,, \quad \omega^n = \dd{r} + W \dd{z} + \bar{W} \dd{\bar{z}} + H \omega^l \,, \quad	\omega^m = \frac{\dd{\bar{z}}}{P \rho} \,,
		\label{eq:solution}
	\end{equation}
	The functions $L$, $W$, $H$, $P$ and $\rho$ are defined by the following relations:
	\begin{align}
		\rho &= -\frac{1}{r + i\Sigma} \,, &2i\Sigma &= P^2 (\Db L - \D\bar{L}) \,,\nonumber\\
		W &= \frac{\partial_u L}{\rho} + \D(i\Sigma) \,,  &\D &= \partial_z - L \partial_u \,, \nonumber\\
		H &= \frac{K}{2} - r \partial_u(\log(P)) - \frac{M r + N \Sigma}{r^2 + \Sigma^2} \,, & K &= 2 P^2 \Re(\D(\Db\log P - \partial_u \bar{L})) \,,\nonumber\\
		N &= \Sigma K + P^2 \Re(\D\Db \Sigma - 2\partial_u\bar{L}\D\Sigma - \Sigma \partial_u\D\bar{L}) \,.
		\label{eq:def-funcs-metric}
	\end{align}
	In this setup, $M$ is a constant which will play the role of the mass while $P(u, z, \bar{z})$ is a real function and $L(u, z, \bar{z})$ is complex. Let us describe the main kinematical properties of the geometries belonging to this family. 
	
	By construction, the geometry being algebraically special, one has $\Psi_0 = \Psi_1 =0$ such that the most general examples are of Petrov type II. The only non-vanishing Weyl scalars are $\Psi_2$, $\Psi_3$ and $\Psi_4$. We do not provide their expressions here as they will not be useful for the rest of our work.
	The Goldberg-Sachs theorem implies that these geometries are equipped with a geodesic and shear-free null congruence such that the Newman-Penrose spin coefficients (NPSC) $\kappa$ and $\sigma$ vanish:
	\begin{align}
		\kappa = \sigma = 0 \,.
	\end{align}
	Except for $\tau$ which also vanishes, all other NPSC are non-vanishing. In particular, one has $\Im(\rho) \neq 0$ which implies that the geometry is twisting. It follows that the null tetrad describes a geodesic, non-shearing but twisting and expanding congruence of null rays. This family of solutions describes the twisting algebraically special vacuum radiative solutions of GR. One can show that under the above assumptions, i.e. choice of tetrad, the Einstein field equations reduce to
	\begin{subequations}
		\begin{align}
			&i \D N - 3(M + iN) \partial_u L = 0 \,,\label{eq:evol-L}\\
			&P\qty[\D + 2(\D\log P - \partial_u\bar{L})] \D\qty[\Db(\Db\log P - \partial_u\bar{L}) + (\Db\log P - \partial_u\bar{L})^2] - \partial_u\qty[P^{-3}(M + i N)] = 0 \,,\label{eq:evol-P}
		\end{align}
	\end{subequations}
	These two equations encode the non-linear dynamical relation between the two functions $P$ and $L$ labelling this solution space. When $L=0$, the twist vanishes and this set of solutions reduces to the Robinson-Trautman family of exact radiative solutions.
	
	The key advantage of working with this solution space is that one can consider a given stationary algebraically special solution belonging to this solution space and linearize the above dynamical equations around that background solution. This will automatically provide a description of the linear ASP of this specific solution. This strategy was first used by Couch and Newman in~\cite{Couch:1973zc} to describe the ASP of the Schwarzschild black hole. This was then revisited by Qi and Schutz in~\cite{Qi:1993ey}, focusing on the non-twisting subsector. Finally, the same approach was used to investigate the quadratic ASP of the Schwarzschild black hole in~\cite{BenAchour:2024skv}. In the following, we shall apply this strategy to the linear ASP of the Kerr black hole. 
	
	
	

	\subsection{Linearization around Kerr: metric perturbations}
	
	Consider hereafter the Kerr solution~\eqref{kerr_metric}. One can show that it corresponds to the couple $(P,L)$ given by
	\begin{equation}
		P = P_0 = \frac{1}{\sqrt{2}} (1 + z\bar{z}) \qq{and} L = L_0 = - \frac{i \bar{z} a}{P_0^2} \,.
	\end{equation}
	Since the metric (\ref{eq:metric}) is written in term of the coordinates $(u, r, z, \bar{z})$, it is useful to provide the explicit coordinate change relating these coordinates to Boyer-Lindquist coordinates $(t, r, \theta, \varphi)$. This mapping reads
	\begin{subequations}
		\begin{align}
			u &= t - r + \frac{2M^2}{\sqrt{M^2 - a^2}} \arctanh\frac{r-M}{\sqrt{M^2 - a^2}} - M \log(r^2 - 2Mr + a^2) \,,\\
			r &= r \,,\\
			z &= \cot\frac{\theta}{2} \exp\qty[i\big(\varphi - \frac{a}{\sqrt{M^2 - a^2}}\arctanh\frac{r-M}{\sqrt{M^2 - a^2}} \big)] \,,\\
			\bar{z} &= \cot\frac{\theta}{2} \exp\qty[-i\big(\varphi - \frac{a}{\sqrt{M^2 - a^2}}\arctanh\frac{r-M}{\sqrt{M^2 - a^2}} \big)] \,.
		\end{align}
		\label{eq:stereo-to-BL}
	\end{subequations}
	Now, the ASP of the Kerr black hole can be described in terms of the perturbed fields $(F_1,L_1)$ which are related to the fields $(P,L)$ through
	\begin{equation}
		P(u, z, \bar{z}) = P_0 e^{\veps F_1} \,,\quad L(u, z, \bar{z}) = L_0 + \veps L_1 \,,
	\end{equation}
	Similarly, we define the expansion of all functions appearing in~\eqref{eq:def-funcs-metric}. For instance, we shall write $N = N_0 + \veps N_1$. 
	Here $\veps$ is a bookkeeping parameter encoding the order of the expansion in perturbation. At this stage, one has to also expand the differential operator defined in (\ref{eq:def-funcs-metric}) as it contains the field $L$. Let us therefore introduce the notation
	\begin{equation}
		\D = \partial_0 + \veps \partial_1 \qq{with} \partial_0 = \partial_z - L_0 \partial_u \qq{and} \partial_1 = -L_1 \partial_u \,.
	\end{equation}
	This expansion implies that we have the following commutation relations
	\begin{equation}
		\comm{\partial_0}{\partial_u} = 0 \,,\quad \comm{\partial_0}{\bar{\partial}_0} = \frac{2i\Sigma_0}{P_0^2} \partial_u \,.
	\end{equation}
	where $\Sigma_0 = a \cos\theta$.
	Finally, it will also reveal useful to introduce the differential operators defined by
	\begin{align}
		\label{opp}
		\Dz &= 2 P_0^2 \Re(\partial_0\bar{\partial}_0 \cdot) \,, &\partial_\varphi &= i \big(z \partial_z - \bar{z} \partial_{\bar{z}} \big) \,, &\delta &= z \partial_z + \bar{z} \partial_{\bar{z}} \,.
	\end{align}
	In term of the BL coordinates, these operators take the form
	\begin{equation}
		\Dz = \Delta_0 - 2 a \partial_u \partial_\varphi + a^2 \sin^2\theta \partial_{uu} \,, \qquad \delta = -\sin\theta \partial_\theta \,,
	\end{equation}
	where $\Delta_0$ is the Laplacian operator on the 2-sphere. Let us point that the operator $\Dz$ is very similar to the operator introduced in~\cite{Giorgi:2021skz}.
	We are now ready to linearize the dynamical equations (\ref{eq:evol-L}) and (\ref{eq:evol-P}) w.r.t the Kerr background.
	
	Consider the first equation~\eqref{eq:evol-L} and let us introduce the field $f_1$ defined by
	\begin{equation}
		L_1 = i \partial_0 f_1 \,.
	\end{equation}
	At first order, this equation leads to 
	\begin{equation}
		\label{GO}
		i \partial_0 N_1 = 3 M \partial_u L_1  \qquad \rightarrow \qquad N_1 = 3 M \partial_u f_1 \,.
	\end{equation}
	Then, using the definiton~\eqref{eq:def-funcs-metric} and the operators~\eqref{opp}, one can compute the first order quantity $N_1$ which reads
	\begin{equation}
		N_1 = \frac14 \Dz\Dz f_1 + \frac12 \Dz f_1 - \Big(3 \Sigma_0^2 - \frac{4 a^2 z \bar{z}}{(1+z\bar{z})^2}\Big) \partial_{u u} f_1  - a \partial_u\partial_\varphi f_1 +2 \Sigma_0 \Dz F_1 + 2 a \delta F_1 \,.
	\end{equation}
	Plugging this expression in (\ref{GO}), one finally obtains
	\begin{equation}
		\boxed{\frac14 \Dz\Dz f_1 + \frac12 \Dz f_1 - \Big(3 \Sigma_0^2 - \frac{4 a^2 z \bar{z}}{(1+z\bar{z})^2}\Big) \partial_{u u} f_1  - a \partial_u\partial_\varphi f_1 - 3M \partial_u f_1 = - 2 \Sigma_0 \Dz F_1 - 2 a \delta F_1 + \mathcal{B}} \,,
		\label{eq:pert-axial}
	\end{equation}
	where $\mathcal{B}$ is a constant. It provides a first coupled PDE for the perturbations $(f_1,F_1)$. When the rotation is turned off, i.e. $a=0$ and thus $\Sigma_0 =0$, the contributions involving $F_1$ vanish and one recovers the equation (3.7) in~\cite{BenAchour:2024skv} which encodes the dynamics of the axial ASP on the Schwarzschild background. Let us point that in \cite{BenAchour:2024skv}, the constant $\mathcal{B}$ was missed (set to zero). As we shall see, it will play a key role when deriving the stationary zero modes of the Kerr black hole.
	
	Similarly, using the same strategy, one can also show that the second dynamical equation~\eqref{eq:evol-P} can be recast in the form
	\begin{equation}
		\boxed{\frac14 \Dz\Dz F_1 + \frac12 \Dz F_1 - \Big(3 \Sigma_0^2 - \frac{4 a^2 z \bar{z}}{(1+z\bar{z})^2}\Big) \partial_{u u} F_1  - a \partial_u\partial_\varphi F_1 + 3M \partial_u F_1 = \partial_{uu} \big(2 \Sigma_0 \Dz f_1 + 2 a \delta f_1 \big)} \,.
		\label{eq:pert-polar}
	\end{equation}
	Setting $a=0$, this equation reduces to the dynamical equation (3.9) of~\cite{BenAchour:2024skv} encoding the polar ASP on the Schwarzschild background. The two above equations provide the first result of this work. They stand as a direct extension to the Kerr background of the equations derived initially by Couch and Newman in~\cite{Couch:1973zc} and later revisited by Qi and Schutz~\cite{Qi:1993ey} for the Schwarzschild black hole background.
	
	As expected, the inclusion of rotation breaks the decoupling of the two perturbations $(f_1,F_1)$. The linear ASP thus provides a simple and explicit example of the complicated coupling that shows up in the metric formulation of the Kerr perturbations. 
	Nevertheless, while solving this system of coupled PDE is full generality is challenging, we will show that one can construct exact analytic solutions in the slow rotating regime. As we shall see, the profiles for the $(f_1,F_1)$ perturbations will reveal interesting features of the metric formulation. To our knowledge, such exact analytic solutions of the ASP have not been presented elsewhere so far.
	
	\section{Dynamical algebraically special modes in the low-spin regime}
	
	\label{III}
	
	In this section, we first present a strategy to solve the system~\eqref{eq:pert-axial} and~\eqref{eq:pert-polar} based on an ansatz for the fields $(f_1,F_1)$. Then, we present one given solution up to third order in the spin parameter $a$ which describes a non-trivial propagating Kerr ASP.
	Such a propagating AS mode satisfies 
	\be
	\delta \Psi_0 = 0 \,, \qquad \delta \Psi_4 \neq 0 \,.
	\ee
	\subsection{Resolution scheme and exact solution}
	
	In order to simplify the resolution of the coupled equations~\eqref{eq:pert-axial} and~\eqref{eq:pert-polar}, we will need to introduce a specific ansatz for the perturbations $(f_1,F_1)$. To that end, it is natural to adopt the same strategy introduced in~\cite{Couch:1973zc,BenAchour:2024skv} for the Schwarzschild ASP and assume that  the $(f_1,F_1)$ profiles can be separated as
	\begin{equation}
		\label{ap}
		f_1(u, z, \bar{z}) = \mathcal{E}_a e^{\kappa u} f(z, \bar{z}) \,,\qquad F_1(u, z, \bar{z}) = \mathcal{E}_p e^{\kappa u} F(z, \bar{z}) \,,
	\end{equation}
	where we have defined $\kappa=i\omega$ for convenience and where $(\mathcal{E}_a, \mathcal{E}_p)$ are constant amplitude factors. The subscript $a$ and $p$ will be explained in a moment.

	Consider now an expansion in the spin parameter $a$ both for the frequency $\omega$ and for the angular waveforms $f(z, \bar{z})$, $F(z, \bar{z})$ such that
	\begin{subequations}
		\begin{align}
			\omega &= \sum \omega_\bn a^\bn \,,\\
			f(z, \bar{z}) &= \sum_{\ell' m' \bn} c_{\ell' m' \bn} Y_{\ell' m'} a^\bn \,,\\
			F(z, \bar{z}) &= \sum_{\ell' m' \bn} d_{\ell' m' \bn} Y_{\ell' m'} a^\bn \,.
		\end{align}
		\label{eq:decompo-a}
	\end{subequations}
	where $\bm{n}$ controls the order to low spin expansion for the different quantities and $Y_{\ell m} $ are the standard spherical harmonics defined by
	\begin{equation}
		Y_{\ell m} = (-1)^m \sqrt{\frac{2\ell+1}{4\pi} \frac{(\ell-m)!}{(\ell+m)!}} P_{\ell m}\Big(\frac{-1+z \bar{z}}{1+z\bar{z}}\Big) \Big(\frac{z}{\bar{z}}\Big)^{m/2} \,,
	\end{equation}
	where $P_{\ell m}$ are Legendre polynomials. While using a decomposition in spherical harmonics for our problem might appears curious at first, it turns out that it is actually the most direct way to separate and thus simplify the two equations~\eqref{eq:pert-axial} and~\eqref{eq:pert-polar}. The key reason is that the angular part of these two dynamical equations has been recast in terms of the operators (\ref{opp}). In particular, the operator $\Delta_0$ is nothing else than the Laplacian operator on the 2-sphere, making the use to the spherical harmonics natural in our case. Nevertheless, it is only a technical trick to solve the above specific equations. Let us further point that despite the fact that $f_1$ and $F_1$ are required to be real functions, we have decomposed them onto complex functions. However, this is not an issue as the real solutions can be recovered by considering the combinations $(f_1 + \bar{f}_1)/2$ and $(F_1 + \bar{F}_1)/2$.
	
	Plugging the above ansatz for $(f_1,F_1)$ in~\eqref{eq:pert-axial} and~\eqref{eq:pert-polar} and solving at each order $\bm{n}$ in the parameter $a$, one then looks for the expressions of the coefficients
	\be
	\label{solcoeff}
	\omega_\bn\,, \qquad c_{\ell' m' \bn} \qq{and} d_{\ell' m' \bn} \,.
	\ee 
	This provides a simple and tractable algorithm to extract the explicit perturbations profiles. Notice that a consistency check of this resolution scheme implies that once the frequency $\omega$ is obtained, one can check that it indeed satisfies the equation (\ref{eq:det-chandra}) obtained by Wald, up to the given order in the expansion. 
	
	Let us now explain the use of the subscript $a$ and $p$ in (\ref{ap}). Consider first the case $a = 0$, corresponding to the Schwarzschild ASP. In the perturbative scheme used here, it corresponds to fixing $\bm{n}=0$. Choosing a given pair of ($\ell$,$m$), one can show that the coefficients of the decomposition are given by
	\begin{equation}
		c_{\ell' m' \bm{0}} = \begin{cases*}
			1 & if $\ell' = \ell$ and $m' = m$ \\0 & else 
		\end{cases*}
		\qq{and}
		d_{\ell' m' \bm{0}} = 0 \,,
		\label{eq:ax-branch}
	\end{equation}
	or
	\begin{equation}
		c_{\ell' m' \bm{0}} = 0 \qq{and}
		d_{\ell' m' \bm{0}} = \begin{cases*}
			1 & if $\ell' = \ell$ and $m' = m$ \\ 0 & else 
		\end{cases*}
		\,.
		\label{eq:pol-branch}
	\end{equation}
	These two distinct cases correspond to the splitting between axial and polar perturbations in the Schwarzschild case. The first one corresponds to a pure axial mode encoded in $f_1$ while the second corresponds to a pure polar mode encoded in $F_1$. This reproduces the solutions derived in~\cite{Couch:1973zc,BenAchour:2024skv}.
	
	Let us now consider the case $a\neq 0$. Since we are interested in solving the perturbations profile in a perturbative scheme, we can use the splitting between axial and polar sectors of the zeroth order (corresponding to the Schwarzschild case) to organize the higher order of the perturbation scheme. In the following, we refer to the first branch~\eqref{eq:ax-branch} as the \textit{axial branch} and the second branch~\eqref{eq:pol-branch} as the \textit{polar branch}, respectively denoted in the following with a superscript \enquote{$\mathrm{a}$} and a superscript \enquote{$\mathrm{p}$}. Thus, the \textit{axial} and \textit{polar} coefficients for $\bm{n}=0$~\eqref{eq:ax-branch} and~\eqref{eq:pol-branch} can then be written as
	\begin{align}
		c^\mathrm{a}_{\ell' m' \bm{0}} &=  \delta_{\ell}^{\ell'} \delta_m^{m'}  &\text{and}&&
		d^\mathrm{a}_{\ell' m' \bm{0}} &= 0
		\,,\\
		c^\mathrm{p}_{\ell' m' \bm{0}} &= 0 &\text{and}&&
		d^\mathrm{p}_{\ell' m' \bm{0}} &= \delta_{\ell}^{\ell'} \delta_m^{m'}
		\,.
		\label{branches}
	\end{align}
	Notice that this splitting is intimately related to the perturbative resolution scheme used here. Let us now present a concrete example of this resolution algorithm and compute the first coefficients of the expansions for a specific case. In the following, we will set $M=1$. 
	
	\subsection{An illustrative example}
	
	\label{exact}
	
	Consider the case where $\ell = 3$, $m = 1$ which was treated for the Schwarzschild case in~\cite{BenAchour:2024skv}. Our goal is to solve the perturbations profiles and the associated frequency up to third order in the spin parameter $a$.
	
	\begin{itemize}
		\item \textit{ASP frequency:} Let us first start with the expansion coefficients of the frequency $\omega$. 
		We find for the axial branch
		\begin{align}
		\label{freqa}
			\omega^\mathrm{a}_{\bm{0}} &= -10i \,,  &\omega^\mathrm{a}_{\bm{1}} &= \frac{100}{3} \,, &\omega^\mathrm{a}_{\bm{2}} &= \frac{4250}{27}i  \,, &\omega^\mathrm{a}_{\bm{3}} &= -\frac{226250}{243} \,,
		\end{align}
		and for the polar branch
		\begin{align}
				\label{freqp}
			\omega^\mathrm{p}_{\bm{0}} &= 10i \,,  &\omega^\mathrm{p}_{\bm{1}} &= \frac{100}{3} \,, &\omega^\mathrm{p}_{\bm{2}} &= -\frac{4250}{27}i  \,, &\omega^\mathrm{p}_{\bm{3}} &= -\frac{226250}{243} \,.
		\end{align}
		As expected, the zeroth order is purely imaginary in agreement with the properties of the Schwarzschild ASP. Then, one can notice that these values correspond to the different orders of expansion~\eqref{expansion_om} of the roots $\omega^\pm$ of the equation~\eqref{eq:det-chandra}, computed for $\ell=3$ and $m=1$. Therefore, our results are in agreement with the famous equation~\eqref{eq:det-chandra} found by Wald which stands as an explicit definition of the propagating Kerr ASP~\cite{Wald:1973wwa}. 
		\item \textit{Coefficients of the axial branch}: Focusing now on the coefficients $c_{\ell' m' \bn}$ and $d_{\ell' m' \bn}$ of the expansions of the angular waveforms $f(z,\bar{z})$ and $F(z,\bar{z})$, we find for the axial branch:
		\begin{subequations}
			\begin{align}
				c^\mathrm{a}_{31\bm{0}} &= 1 \,, &c^\mathrm{a}_{11\bm{2}} &= \frac{20}{9}\sqrt{\frac27} \,, &c^\mathrm{a}_{51\bm{2}} &= -\frac{20}{3} \sqrt{\frac{10}{77}} \,, &c^\mathrm{a}_{11\bm{3}} &= \frac{3200}{81} i \sqrt{\frac27} \,, \\
				c^\mathrm{a}_{51\bm{3}} &= -\frac{100}{27} i \sqrt{\frac{110}{7}}\,, &d^\mathrm{a}_{21\bm{1}} &= -\frac{160}{9}\sqrt{\frac{10}{7}} \,, &d^\mathrm{a}_{41\bm{1}} &= -25\sqrt{\frac{5}{21}} \,, &d^\mathrm{a}_{21\bm{2}} &= -\frac{7600}{81} i \sqrt{\frac{10}{7}} \,,\\
				d^\mathrm{a}_{41\bm{2}} &= -\frac{425}{3}i \sqrt{\frac{5}{21}} \,, &d^\mathrm{a}_{21\bm{3}} &= \frac{337100}{729} \sqrt{\frac{10}{7}} \,, &d^\mathrm{a}_{41\bm{3}} &= \frac{1026125}{1188} \sqrt{\frac{5}{21}} \,, &d^\mathrm{a}_{61\bm{3}} &= \frac{15500}{297} \sqrt{\frac{2}{13}} \,.
			\end{align}
		\end{subequations}
		The values of $c^\mathrm{a}_{\ell'm'\bn}$ and $d^\mathrm{a}_{\ell'm'\bn}$ which are not mentioned vanish.
		\item \textit{Coefficients of the polar branch}: The coefficients of the polar branch are given by
		\begin{subequations}
			\begin{align}
				c^\mathrm{p}_{21\bm{1}} &= \frac89 \sqrt{\frac{2}{35}} \,, &c^\mathrm{p}_{41\bm{1}} &= \frac14 \sqrt{\frac{5}{21}} \,, &c^\mathrm{p}_{21\bm{2}} &= \frac{20}{81} i \sqrt{\frac{10}{7}}\,, &c^\mathrm{p}_{41\bm{2}} &= \frac14 i \sqrt{\frac{5}{21}}\,,\\
				c^\mathrm{p}_{21\bm{3}} &= \frac{949}{729} \sqrt{\frac{10}{7}}\,,&c^\mathrm{p}_{41\bm{3}} &= \frac{545}{1584} \sqrt{\frac{5}{21}} \,,&c^\mathrm{p}_{61\bm{3}} &= -\frac{155}{297} \sqrt{\frac{2}{13}}\,, &d^\mathrm{p}_{31\bm{0}} &= 1 \,,\\
				d^\mathrm{p}_{11\bm{2}} &= \frac{20}{9}\sqrt{\frac27} \,, &d^\mathrm{p}_{51\bm{2}} &= -\frac{20}{3} \sqrt{\frac{10}{77}} \,, &d^\mathrm{p}_{11\bm{3}} &= -\frac{3200}{81} i \sqrt{\frac27} \,, & d^\mathrm{p}_{51\bm{3}} &= \frac{100}{27} i \sqrt{\frac{110}{7}}\,.
			\end{align}
		\end{subequations}
		The values of $c^\mathrm{p}_{\ell'm'\bn}$ and $d^\mathrm{p}_{\ell'm'\bn}$ which are not mentioned vanish. 
	\end{itemize}
	The above example provides a first exact analytic solution of the Kerr ASP profiles up to third order in the spin which satisfies the Wald condition (\ref{eq:det-chandra}). We stress that this exact solution is not generic and has been obtained by imposing a suitable ansatz. Scanning the whole set of possible solutions would require a more detailed exploration. However, it is worth emphasizing that as they stand, the coupled dynamical equations are still highly complicated to solve analytically.
	
	At this stage, there are three interesting observations to make on the analytic profiles derived above. First, in contrast to the Schwarzschild ASP, the $(F_1,f_1)$ profiles are not purely growing or purely decaying, but inherit a mixing between both. Second, we observe that in contrast to the Weyl perturbations encoded in the Teukolsky equation, the decomposition of angular part of the metric perturbations is not related to the spin-weighted spheroidal harmonics. Finally, notice that as anticipated, and in contrast to their Schwarzschild counterpart, the Kerr ASP admit an oscillating contribution. Indeed, in the case $a = 0$, one recovers the profiles of the Schwarzschild ASP given by
	\begin{equation}
		f_1 = E_\mathrm{a} Y_{\ell_\mathrm{a}m_\mathrm{a}} e^{\kappa_\mathrm{a} u} \,,\qquad F_1 = E_\mathrm{p} Y_{\ell_\mathrm{p}m_\mathrm{p}} e^{-\kappa_\mathrm{p} u} \,,
	\end{equation}
	with
	\begin{equation}
		\kappa(\ell) = \frac{(\ell-1)\ell(\ell+1)(\ell+2)}{12M} \,.
	\end{equation}
	In the present example, where $\ell = 3$, $m = 1$, one obtains $\kappa = 10$, in agreement with the zeroth order of (\ref{freqa}) and (\ref{freqp}). Having describe how to solve for the profiles of a subsector of the dynamical Kerr ASP, we now turn to the discussion of the zero modes.
	
	\section{Stationary zero modes}
	
	\label{IV}
	
	In this section, we focus on the stationary zero modes of the Kerr black hole, i.e. on the non-radiative algebraically special modes. By construction, these zero modes fall in the solution space described in Section~\ref{II}. As discussed in Section~\ref{Wald}, it was shown by Wald that the stationary zero modes form a four dimensional solution space of the linearized Einstein equations. The first two solutions correspond to a Kerr black hole with a perturbation in mass or spin. The third one is the linearized Kerr-NUT solution while the fourth one corresponds to the linearized spinning C-metric. Each of these solutions is of Petrov type D such that they satisfy
	\be
	\delta \Psi_0 = \delta \Psi_4 =0 \,.
	\ee
	Our goal here is to derive the analytic expression for the perturbations relating the Kerr solution to each of these four exact solutions.
	
	\subsection{Perturbations in mass and spin}
	
	Let us start with the perturbations corresponding to a shift of mass and spin. For the Schwarzschild black hole, it is well known that the $\ell=0$ and $\ell=1$ perturbations are either pure gauge (they do not induce any physical change on the background charge), or are large gauge transformations inducing a physical shift in the black hole charges. For instance, the $\ell=1$ axial perturbation turns the static Schwarzschild solution into the slowly rotating Kerr solution, effectively inducing a non-vanishing spin to the background. In the following, we derive the explicit expressions of the polar and axial perturbations of the Kerr black hole and discuss their respectively small or large gauge characters.
	
	\subsubsection{Polar and axial monopole perturbation}
	
	Let us start with the monopole perturbations of the Kerr black hole. We define the axial and polar monopole perturbations from their Schwarzschild limit:
	\begin{align}
		& \text{axial monopole}: & c^\mathrm{a}_{\ell' m' \bm{0}} &=  \delta_{0}^{\ell'} \delta_0^{m'} \,, & d^\mathrm{a}_{\ell' m' \bm{0}} &= 0 \,,\\
		& \text{polar monopole}: & c^\mathrm{p}_{\ell' m' \bm{0}} &=  0 \,, &d^\mathrm{p}_{\ell' m' \bm{0}} &= \delta_{0}^{\ell'} \delta_0^{m'} \,.
	\end{align} 
	In terms of perturbation profiles, monopole perturbations can be written as
	\begin{align}
	\label{mona}
		& \text{axial monopole}: & f_1 &= \mathcal{E}_{\mathrm{a}} Y_{00} \,, & F_1 &= 0 \,,\\
			\label{monp}
		& \text{polar monopole}: & f_1 &= 0 \,, & F_1 &= \mathcal{E}_{\mathrm{p}} Y_{00} \,,
	\end{align} 
	which are both trivial solutions of equations~\eqref{eq:pert-axial} and~\eqref{eq:pert-polar}. Using these profiles, one can then evaluate the effect of such perturbations on the metric~\eqref{eq:metric}. 
	
	One finds that the presence of an axial monopole perturbation leaves $\dd{s}^2$ unchanged while a polar monopole perturbation induces \textit{both} a shift of mass, i.e. $M + \varepsilon \var{M}$, and spin, i.e. $a + \varepsilon \var{a}$ of the form
	\begin{equation}
	\label{shiftmon}
		\var{M} = - \frac{3M}{2\sqrt{\pi}} \mathcal{E}_\mathrm{p} \,,\qquad \var{a} = \frac{a}{2\sqrt{\pi}} \mathcal{E}_\mathrm{p} \,.
	\end{equation}
	The shift of mass and spin is not direct to determine. The perturbed metric $\bar{g}_{\mu\nu}+ h_{\mu\nu}$ dressed with a monopole polar perturbation $h_{\mu\nu}$ takes a complicated form with non-trivial off diagonal terms. To read the change in mass and spin, one has to bring back this perturbed metric to the Kerr metric in Boyer-Lindquist coordinates to identify both $\delta M$ and $\delta a$. At first order in $\epsilon$, this change of coordinates is explicitly given by 
	\begin{subequations}
	\label{CCH}
		\begin{align}
			t &= \!\begin{aligned}[t]&\Big(1 - \varepsilon \frac{\mathcal{E}^\mathrm{p}}{2\sqrt{\pi}}\Big) T + \frac{\varepsilon\mathcal{E}^\mathrm{p}}{\sqrt{\pi}} \Bigg[r + \frac{2 M \left(a^4+a^2 M^2-2 M^3 r\right)}{\left(M^2-a^2\right) \left(a^2-2 M r+r^2\right)} \\& - 4M^2 \frac{M^2 - 2a^2}{(M^2 -a^2)^{3/2}} \arctanh\frac{r-M}{\sqrt{M^2 - a^2}} + 2M \log(r^2 -2MR + a^2)\Bigg] \,,\end{aligned}\\
			r &= \Big(1 + \varepsilon \frac{\mathcal{E}^\mathrm{p}}{2\sqrt{\pi}}\Big) R  \,,\\
			\theta &= \Theta \,,\\
			\varphi &= \Phi + \frac{a \varepsilon \mathcal{E}^\mathrm{p}}{\sqrt{\pi}} \qty[\frac{a^2 (r-2 M)+M^2 r}{\left(M^2-a^2\right) \left(a^2+r (r-2 M)\right)} - \frac{2M^2}{(M^2 - a^2)^{3/2}} \arctanh\frac{r-M}{\sqrt{M^2 - a^2}}] \,.
		\end{align}
	\end{subequations}
	Therefore, while the axial monopole perturbation is a pure gauge, the polar monopole perturbation stands as large gauge transformation. Notice that the situation contrasts with the effect of a polar monopole perturbation on the Schwarzschild background which only induces a shift of mass. We can now consider the effects of a dipole perturbation.
	
	\subsubsection{Axial and polar dipole perturbations}
	
	Similarly to the monopole, we define dipolar perturbations as the $a \neq 0$ extension of the Schwarzschild case. We therefore consider axial and polar dipoles with $\ell = 1$ and $m \in \{-1, 0, 1\}$. The cases $m = -1$ and $m = 1$ are similar so we only consider $m = 1$. 
	
	Let us first consider the case of the axial dipole perturbations. For $m = 0$, the axial dipole is given by
	\begin{equation}
	\label{dipa}
		f_1 = \mathcal{E}_{\mathrm{a}} Y_{10}  \qq{and} F_1 = 0 \,.
	\end{equation}
	Plugging this perturbation in the metric, one finds that the metric $\bar{g}_{\mu\nu} + h_{\mu\nu}$ corresponds to a Kerr black hole with a change in spin $\var{a}$ given explicitly by
	\begin{equation}
	\label{shiftdip}
		\var{a} = \frac12 \sqrt{\frac{3}{\pi}} \mathcal{E}_\mathrm{a} \,.
	\end{equation}
	Identifying this shift in spin requires the following change of coordinates from $(t, r, \theta, \varphi)$:
	\begin{subequations}
		\begin{align}
			t &= T -2 \varepsilon a M \sqrt{\frac{3}{\pi}} \mathcal{E}^\mathrm{a} \qty[\frac{M r-a^2}{2 \left(a^2-M^2\right) \left(a^2-2 M r+r^2\right)} + \frac{M}{2 \left(M^2-a^2\right)^{3/2}} \arctanh\frac{r-M}{\sqrt{M^2 - a^2}}]\,,\\
			r &= R  \,,\\
			\theta &= \Theta \,,\\
			\varphi &= \Phi + \frac{\varepsilon}{2} \sqrt{\frac{3}{\pi}} \mathcal{E}^\mathrm{a} \qty[\frac{a^2 (r-M)}{\left(a^2-M^2\right) \left(a^2+r (r-2 M)\right)} + \frac{M^2}{\left(M^2-a^2\right)^{3/2}} \arctanh\frac{r-M}{\sqrt{M^2 - a^2}} ] \,.
		\end{align}
	\end{subequations}
	Therefore, just as for the Schwarzschild black hole, the axial dipole perturbation stands as a large gauge transformation shifting the spin of the Kerr background.
	
	Let us now focus on the polar dipole. Contrary to the Schwarzschild case, one finds that it involves a mixing between the dipole and quadrupole such that the perturbations $(f_1,F_1)$ are given by
	\begin{equation}
		F_1 = \mathcal{E}_{\mathrm{p}} Y_{10} \qq{and} f_1 = \frac{2a}{\sqrt{15}} \mathcal{E}_{\mathrm{p}} Y_{20} \,.
	\end{equation}
	This decomposition of the polar dipole also illustrates that despite being a technical trick, the decomposition in spherical harmonics is actually not the one adapted to capture the angular profile of these perturbations of the Kerr black hole.
	The resulting perturbed spacetime is related to the unperturbed Kerr black hole via a simple change of coordinates involving only the polar coordinate $\theta$ and given explicitly by
	\begin{equation}
		\theta = \Theta + \frac12 \sqrt{\frac{3}{\pi}} \mathcal{E}^\mathrm{p} \sin\theta \,.
	\end{equation}
	Interestingly, this is exactly what was found in the non-rotating case in~\cite{BenAchour:2024skv}\footnote{For $m = 1$, the axial dipole is given by
		\begin{equation}
			f_1 = \mathcal{E}^{\mathrm{a}} Y_{11}  \qq{and} F_1 = 0 \,, 
		\end{equation}
		while the polar dipole is
		\begin{equation}
			F_1 = \mathcal{E}^{\mathrm{p}} Y_{11} \qq{and} f_1 = \frac{a}{\sqrt{5}} \mathcal{E}^{\mathrm{p}} Y_{21} \,.
	\end{equation}}. It follows that the polar dipole stands as a pure gauge transformation for the Kerr black hole. This concludes the classification of the zero modes of the Kerr black hole involving a shift of the mass and spin. We now turn to the remaining stationary zero modes associated to the Kerr-NUT and spinning C-metric.
	
	\subsection{Kerr-NUT}
	
	It is well-known that the Kerr solution belongs to a larger family of exact vacuum solutions of Einstein equations which are all of Petrov type D, the so called Plebianski-Demianski family~\cite{Plebanski:1976gy, Griffiths:2005qp}. This family is labelled by five parameters which are the mass, the spin, the acceleration, the so called NUT parameter, the electric and magnetic charges.  The Kerr-NUT solution was derived initially by Demianski and Newman in~\cite{Dem}, as an rotating extension of the solution first found by Newman, Tamborino and Unti in~\cite{Newman:1963yy}. This solution is not asymptotically flat and possesses exotic properties which have challenged its interpretation\footnote{In particular, it enjoys a rather surprising duality which allows one to switch the mass and NUT charges together with a switch of the radial and angular coordinates.}. The NUT charge $n$ has been interpreted as a kind of dual mass charge, in the similar sense as the dual role played by the magnetic charge w.r.t the electric charge in Maxwell theory. Physically, the NUT charge induces string-like defects which are called Misner strings~\cite{Misner, Bonnor}. Effects on the geodesic motion have been studied in detail in~\cite{Clement:2015cxa}, suggesting that anticipated pathological behaviors related to these defects are actually absent (see~\cite{Frodden:2021ces} for a recent study of the Kerr-NUT thermodynamics). Being an exact solution of the Einstein equations which corresponds to a deformation of the Kerr geometry, it appears natural to look for the ASP which generate the NUT charge.
	
	In Boyer-Lindquist coordinates, the Kerr-NUT solution is given by
	\begin{equation}
		\begin{aligned}
			\dd{s}^2 &= - \frac{\Delta}{\Sigma} \big[\dd{t} +(2n \cos\theta + 2Cn - a \sin^2\theta)\dd{\varphi}\big]^2 + \frac{\Sigma}{\Delta}\dd{r}^2 +  \Sigma \dd{\theta}^2   \\
			& \;\; + \frac{\sin^2\theta}{\Sigma} \big[a \dd{t} - (r^2 + a^2 + n^2 - 2anC) \dd{\varphi}]^2 \,,
		\end{aligned}
	\end{equation}
	where
	\be
	\Sigma = r^2 + (n + a \cos\theta)^2 \,,\qquad \Delta = r^2 - 2MR + a^2 - n^2 \,.
	\ee
	The NUT charge is denoted $n$ and $C \in \{-1, 0, 1\}$. It is useful to switch to stereographic coordinates $(U, R ,Z, \bar{Z})$ by setting
	\begin{subequations}
		\begin{align}
			t &= U + R - \frac{2M^2}{\sqrt{M^2 - a^2}} \arctanh\Big(\frac{R-M}{\sqrt{M^2 - a^2}}\Big) + M \log(R^2 - 2MR + a^2) \,,\\
			r &= R \,,\\
			\theta &= 2 \arccot\sqrt{Z \bar{Z}} \,,\\
			\varphi &= \frac{1}{2i} \log\frac{Z}{\bar{Z}} - \frac{a}{\sqrt{M^2 - a^2}}\arctanh\Big(\frac{R-M}{\sqrt{M^2 - a^2}}\Big) \,.
		\end{align}
	\end{subequations}
	Then, after a lengthy computation, one can show that the linearized version of this solution for small $n$ can be reproduced as an exact solution to the perturbation equations~\eqref{eq:pert-axial} and~\eqref{eq:pert-polar}. It corresponds to the perturbation profiles given by
	\begin{subequations}
	\begin{align}
		\label{NUT1}
		f_1 &= \log\frac{z^{1-C}}{(1+z \bar{z})^2} \,,\\
		\label{NUT2}
		F_1 &= 0 \,.
	\end{align}
	\end{subequations}
	These profiles solve the equations~\eqref{eq:pert-axial} and~\eqref{eq:pert-polar} for the specific value of the free constant $\mathcal{B}= -1$.
	Equipped with this solution, one can show that the perturbed metric $\bar{g}_{\mu\nu} + h_{\mu\nu}$ can be mapped to the Kerr-NUT solution~\eqref{eq:solution} at first order in $\varepsilon$ provided $\varepsilon = n$ by using the following change of coordinates:
	\begin{align}
		U &= u - \frac{a C}{\sqrt{M^2 - a^2}} \log\frac{r - M + \sqrt{M^2 - a^2}}{r - M - \sqrt{M^2 - a^2}}\,,\\
		R &= r \,,\\
		Z &= z \,.
	\end{align}
	To our knowledge, this provides the first analytic expression for the algebraically special perturbation of the Kerr geometry generating the NUT charge. Notice that this could not have been achieved using solely the Teukolsky formulation. However, it would be interesting to see if this result can be recovered by using the metric reconstruction procedure outlined in~\cite{Berens:2024czo}. We now turn to the last stationary zero mode to determine, namely the one generating the spinning C-metric.
	
	\subsection{Spinning C metric}
	
	The non-spinning C-metric was initially derived by Weyl in~\cite{Weyl}. It describes a system of two Schwarzschild black holes with equal mass and uniformly accelerating away from each other~\cite{C1,C2}. The acceleration is controlled by a new constant parameter on top of the mass. This acceleration can be understood as originating from  the presence of strings which manifest through conical singularity in between the two black holes. See~\cite{Griffiths:2006tk} for a recent account on the interpretation of this exact solution and \cite{Hong:2003gx} for a recent new system of coordinates for the non-rotating case. The spinning extension of the C-metric was obtained by Plebanski and Demianski in~\cite{Plebanski:1976gy}. A discussion on the physical interpretation of the spinning version can be found in \cite{Pravda:2002kj}. Recently, a new coordinate system was introduced by Hong and Teo in~\cite{Hong:2004dm}, providing several useful simplifications to study this spinning solution. In the following, we shall use this later proposal to identify the ASP generating the spinning C-metric.
	
	In the new coordinates introduced in~\cite{Hong:2004dm}, the spinning C-metric is given by
	\begin{multline}
		\dd{s}^2 = \frac{1}{A^2 (x - y)^2} \Big\{\frac{G(y)}{1 + (a A x y)^2} \big[\dd T + a A (1-x^2) \dd\varphi\big]^2 - \frac{1 + (a A x y)^2}{G(y)} (\dd{y})^2 + \frac{1 + (a A x y)^2}{G(x)} (\dd{x})^2 \\+ \frac{G(x)}{1 + (a A x y)^2} \big[(1+ a^2 A^2 y^2)\dd{\varphi} + a A y^2 \dd{T}\big]^2\Big\} \,,
	\end{multline}
	where $A$ is the acceleration parameter. The change to the standard Boyer-Lindquist coordinates $(t, r, \theta, \varphi)$ is simply given by
	\begin{equation}
		T = A t \,,\qquad x = \cos\theta \,,\qquad y = - \frac{1}{Ar} \,.
	\end{equation}
	In order to transform to the stereographic coordinates $(U, R, Z, \bar{Z})$ used in this work, one has to consider instead the following change of coordinates
	\begin{subequations}
		\begin{align}
			t &= U + R - \frac{2M^2}{\sqrt{M^2 - a^2}} \arctanh\Big(\frac{R-M}{\sqrt{M^2 - a^2}}\Big) + M \log(R^2 - 2MR + a^2) \,,\\
			r &= R \,,\\
			\theta &= 2 \arccot\sqrt{Z \bar{Z}} \,,\\
			\varphi &= \frac{1}{2i} \log\frac{Z}{\bar{Z}} + \frac{a}{\sqrt{M^2 - a^2}}\arctanh\Big(\frac{R-M}{\sqrt{M^2 - a^2}}\Big) \,.
		\end{align}
	\end{subequations}
	A lengthy computation reveals that this solution can be generated by the following non-trivial perturbations
	\begin{align}
		\label{C1}
		f_1 &= \frac{4a z \bar{z}}{(1+z\bar{z})^2} \big[M \log(z\bar{z})  -M - u - 2M \log(1+z\bar{z}) \big] \,,\\
		\label{C2}
		F_1 &= \frac{z\bar{z}-1}{1+z\bar{z}} u - \frac{z\bar{z}-1}{1+z\bar{z}} \big[ 1 + \log(z\bar{z}) -2 \log(1 + z \bar{z}) \big] M\,.
	\end{align}
	These perturbations solve the dynamical equations~\eqref{eq:pert-axial} and~\eqref{eq:pert-polar} for the specific value of the free constant $\mathcal{B} = 4 a M$. The spinning C-metric solution~\eqref{eq:solution} can then be recovered at first order in $\varepsilon$ provided $\varepsilon = A$ after performing the following change of coordinates:
	\begin{align}
		U &= u \,,\\
		\label{CCC2}
		R &= r -  \frac{z \bar{z}  -1 }{1 + z \bar{z}}  A r^2\,,\\
		\label{CCC3}
		Z &= z + A z \big[M \log(z\bar{z}) - 2M \log(1 + z\bar{z}) - u\big] \,.
	\end{align}
	This provides the exact analytic form of the ASP generating the linearized spinning C-metric from the Kerr background solution.
	
	\section{Discussion}
	
	\label{V}
	
	In this work, we have investigated the description of the linear algebraically special perturbations of the Kerr black hole and we have shown how they can be described within the metric formulation. This is in contrast with the standard Kerr perturbations for which an analytic control requires a treatment based on the Weyl scalars and the associated Teukolsky master equations. The main reason allowing this pure metric treatment of the linear Kerr ASP is that by construction they fall in the class of the most general algebraically special twisting vacuum solutions to the Einstein equations which are therefore of Petrov type II. Using this solution space for the most general Petrov type II twisting vacuum geometries presented in Section~\ref{II}, one can simply linearize the reduced Einstein equations around the Kerr background to derive the dynamics of the Kerr ASP. These equations are given explicitly in \eqref{eq:pert-axial} and~\eqref{eq:pert-polar} and represent the first result of this work. They stand as the generalization of the equations found by Couch and Newman in  \cite{Couch:1973zc} and rederived later by Qi and Schutz in \cite{Qi:1993ey} for the Schwarzschild black hole. As expected, the resulting PDEs turn out to be coupled making the solution space challenging to scan. Nevertheless, the solutions of this set of wave equations can be split in two groups. The first one consists in dynamical ASP modes which propagate on the Kerr geometry. The second consists in the so called stationary zero modes, the ASP generating the known exact stationary deformations of the Kerr solution which are of Petrov type D. They correspond to the four possible perturbations inducing a shift of mass, spin, uniform acceleration and NUT charge. Let us review the different results obtained in this work regarding both types of ASP.
\begin{itemize}
\item \textit{Resolution scheme for the dynamical ASP}: Focusing first on the dynamical ASP, we have presented in Section~\ref{III} a perturbative resolution algorithm adapted for the slowly rotating regime. The ability to solve analytically this complicated set of coupled wave equations boils down to two combined features. First, despite the breakdown of spherical symmetry, we have shown that the angular part of the wave equations can be written in term of the operators (\ref{opp}) which involves the Laplacian on the $2$-sphere. Then, by using the specific separable ansatz (\ref{ap}) for the perturbations $(f_1,F_1)$ which separates their angular contributions from their radial and temporal ones, these specific modes can be expanded onto standard spherical harmonics which dramatically simplifies the problem. Moreover, since our approach is perturbative in the spin parameter, the zeroth order of the expansion enjoys the standard splitting of the spherically symmetric Schwarzschild case. This splitting can be further extended to the higher order of the expansion, although it does not rely anymore on the properties of the fields under parity transformations which are no longer defined. This pushed us to introduce the notion of \textit{axial and polar branches}. Within this scheme, deriving an exact solution consists in determining i) the ASP frequency and ii) the decomposition coefficients of the $(f_1,F_1)$ perturbations profiles given by (\ref{solcoeff}).
\item \textit{A first exact analytical solution for the dynamical ASP}: Using this resolution scheme, we have derived in Section~\ref{exact} the first exact analytic solution for a dynamical Kerr ASP. It corresponds to the special case $\ell=3$ and $m=1$ w.r.t to the decomposition in spherical harmonics. To demonstrate the power of the method, we have pushed the expansion up to third order in the spin parameter, but we stress that going beyond that order does not lead to any difficulties. Of course, the solution presented here is only one specific example which can be captured with our perturbative scheme. A detailed scan of the solution space of the Kerr ASP would require more work, but it is worth stressing that to our knowledge, no other exact solution of this type has been obtained so far. As a consistency check, we have shown that the ASP frequencies (\ref{freqa} -- \ref{freqp}) we have found up to third order in spin satisfies the general Wald condition (\ref{eq:det-chandra}) defining the Kerr ASP \cite{Wald:1973wwa}, confirming the robustness of our approach. It also demonstrates the power of our metric-based approach as such exact solutions of the ASP could not have been obtained following Wald approach based on the Teukolsky equation and the Newman-Penrose formalism. It would be interesting to see how one can recover this solution using the metric reconstruction algorithm discussed recently in \cite{Berens:2024czo}.
\item \textit{Zero modes shifting the mass and spin}: The metric-based approach adopted here also allowed us to present a detailed study of the stationary zero modes of the Kerr black hole perturbations. Interestingly, the decomposition of the ASP in terms of \textit{axial and polar branches} and the decomposition in spherical harmonics turned out to be useful in this task. First, we have shown that the monopole perturbation belonging to the axial branch given by (\ref{mona}) is a pure gauge. On the contrary, the monopole perturbation belonging to the polar branch (\ref{monp}) was shown to generate both a shift of mass and spin given explicitly by (\ref{shiftmon}). At the dipole level, the polar perturbation was found to be a pure gauge while the axial perturbation (\ref{dipa}) stands as a large gauge transformation inducing a shift in spin given by (\ref{shiftdip}), just as its counterpart in the spherically symmetric context. While these results have been known for a long time at the qualitative level, we have provided an explicit derivation of the associated perturbations profiles and more importantly, we have provided the explicit expressions for the shifts in mass and spin w.r.t to the standard Boyer-Lindquist coordinates. It is worth emphasizing that identifying these shifts is a non trivial task as it requires finding suitable changes of coordinates which can be especially involved, see for instance (\ref{CCH}).
\item \textit{Zero modes profiles generating the NUT charge and acceleration} : We have also derived the analytic expression for the ASP generating the two other non-trivial stationary deformations of the Kerr solution, namely the NUT and the acceleration charges. First we have shown that the perturbed Kerr black hole metric with the perturbation profile (\ref{NUT1}) can be transformed into the Kerr-NUT metric linearized w.r.t the NUT charge. We stress that this task is highly non-trivial as it also requires identifying the required change of coordinates given by (\ref{C1} -- \ref{C2}). Similarly, we have shown that using the perturbation (\ref{C1} -- \ref{C2}) and the change of coordinates (\ref{CCC2} -- \ref{CCC3}), one recovers the spinning C-metric linearized w.r.t the acceleration parameter $A$. To our knowledge, this provides the first derivation of the explicit analytic expressions for the ASP generating these two deformations of the Kerr solution. Let us stress that the role of the constant $\mathcal{B}$ entering in the wave  equation (\ref{eq:pert-polar}) is crucial to derive these results. Indeed, this constant was missed and set to zero in \cite{BenAchour:2024skv}, which explains why the NUT and acceleration charges were not found in our previous work on the Schwarzschild ASP.
\end{itemize}
Having reviewed the main points of this work, let us make some comments on the open directions. Regarding the zero modes, a rigorous classification of the small versus large gauge nature of the different perturbations would require computing the boundary charges associated to the diffeomorphisms generating each of these ASP. While the small gauge perturbations will have vanishing boundary charges, the large gauge ones will induce a non-vanishing contribution on the boundary. This rigorous computation will be presented in an upcoming work.

Another interesting generalization of the present results would be to study the Kerr ASP using the solution space for the most general algebraically special twisting electro-vacuum solutions of GR. This would allow one to capture additional zero modes connecting the Kerr solution to all the representants of the well-known Plebianski-Demianski family \cite{Plebanski:1976gy, Griffiths:2005qp}. An interesting question which naturally arises from this context is whether one can use the approach presented here to explore the existence of hidden symmetries for black hole perturbations. By definition, the ASP enjoy a higher degree of symmetry than the general Kerr perturbations and they represent therefore an interesting starting point to investigate this question.

Finally, let us stress that the present work could also reveal useful to study the non-linear Kerr perturbations, a highly challenging topic so far. The non-linear regime of the black hole ringdown has attracted much attention in the last years \cite{London:2014cma, Cheung:2022rbm, Mitman:2022qdl, Ma:2022wpv, Lagos:2022otp, Bhagwat:2023fid, Kehagias:2023ctr, Khera:2023oyf, Redondo-Yuste:2023seq, Perrone:2023jzq, Bucciotti:2023ets, Cheung:2023vki, Qiu:2023lwo, Zhu:2024rej, Yi:2024elj, Bucciotti:2024zyp, Bourg:2024jme, DeLuca:2023mio, Kehagias:2024sgh, Kehagias:2025ntm, Kehagias:2025xzm, Kehagias:2025tqi, Perrone:2025zhy, Ma:2024qcv,Khera:2024bjs}. In \cite{BenAchour:2024skv}, we have shown that the same approach used here can allow one to analytically compute the quadratic ASP profiles of the Schwarzschild black hole. Extending this result to the Kerr ASP is in principle possible, although quite heavy to realize in practice. See \cite{Ma:2024qcv,Khera:2024bjs} for recent works on the quadratic perturbations of the Kerr black hole.

	\bibliographystyle{JHEP}
	\bibliography{biblio}

\end{document}